\documentclass[aps,prb,preprint,superscriptaddress]{revtex4}

\usepackage{epsfig} 

\begin{document}

\title{Macroscopic Phase Coherence of 
Defective Vortex Lattices in Two Dimensions}
\author{J. P. Rodriguez}
\affiliation{Department of Physics and Astronomy, California State University,
Los Angeles, California 90032, USA.}

\date{\today}

\begin{abstract}
The superfluid density is calculated theoretically for
incompressible vortex lattices in two dimensions that have 
isolated dislocations quenched in 
by a random arrangement of pinned vortices.
The latter
 are  assumed to be sparse and to be  fixed
to material defects.
It is shown that the pinned vortices act to  
confine a  single dislocation of the vortex lattice along its  glide plane.
Plastic creep of the two-dimensional vortex lattice
is thereby impeded,
and macroscopic phase coherence
results at low temperature
in the limit of a dilute concentration of quenched-in dislocations.
\end{abstract}

\maketitle

\section{Introduction}
\label{intro}

Consider either  a bismuth-based, a   mercury-based, or
a  thallium-based high-temperature superconductor in
high enough  external magnetic field
so that magnetic flux lines appear,
and so that these overlap considerably.
Such  materials are extreme  type-II layered superconductors, 
with   mass-anisotropy
ratios between the layer direction and the perpendicular direction
of order $10^{-3}$ or less\cite{blatter}.
To a first approximation then, 
it becomes valid to neglect
the coupling of magnetic screening currents between layers,
as well as the Josephson effect   between them.
The description of the initial physical situation
is thereby reduced
to a stack of isolated vortex lattices or
vortex liquids  within each layer.

High-temperature superconductors like those just mentioned typically
also have crystalline defects and inhomogeneities that act as pinning
centers for vortex lines in the mixed phase.
Within the preceding approximation,
theoretical and numerical studies indicate that {\it any} 
net concentration of
randomly pinned  vortices results in a net concentration of 
unbound dislocations  quenched into the
two-dimensional (2D) vortex lattices
of  each 
layer\cite{M-E}$^,$\cite{berlinsky}$^,$\cite{N-S}$^,$\cite{zeng}.
Defective vortex matter is then left as the only possible
solid state of the mixed phase in two dimensions.
It is useful to divide the  last group into two classes:
({\it i}) 
configurations of vortices
that contain no unbound disclinations,
and ({\it ii}) 
configurations of vortices  
that contain some concentration  of  unbound disclinations\cite{N-H}.  
The amorphous vortex glass characterized by macroscopic phase coherence
falls into the second class\cite{ffh}.
It is believed to exist only at zero temperature in two dimensions, however.
Defective vortex lattices with no unbound disclinations,
but with isolated dislocations\cite{jpr04a},
or with dislocations arranged into 
grain boundaries\cite{chandran}$^,$\cite{menghini}$^,$\cite{moretti},
are  then perhaps  left  as the only 
solid  states of the mixed phase
that are possible in two dimensions above zero temperature.

In this paper, we demonstrate theoretically 
that defective vortex lattices in two
dimensions show macroscopic phase coherence in the extreme
type-II limit, in the regime of weak random pinning.
We find, in particular, that
the 2D vortex lattice  exhibits   a net superfluid density 
if it is  void of disclinations, and if only
a small number of isolated dislocations are quenched in
in comparison to the total number of pinned vortices.
This result is achieved in three steps. 
First, we demonstrate in section II that  a   network of pinned vortices
confines  the motion of a single dislocation along its glide  plane.
This guarantees that the 2D vortex lattice remains elastic in the
limit of a dilute concentration of such unbound dislocations.
Next, the uniformly frustrated $XY$ model for
the  2D vortex lattice is introduced in section III, through which
a useful expression for the superfluid density is derived
in terms of glide by unbound dislocations\cite{jpr04a}.  
Interactions among the dislocations are notably ignored here.
These results are then assembled in section IV, where the
final formula [Eq. (\ref{rho2d4})] for the superfluid density of the
defective  vortex lattice  is obtained
as a function of the ratio of 
the number of unbound dislocations to the number of pinned vortices.

\section{Collective Pinning of One Dislocation}
\label{1df}
It is  strongly
believed that the  vortex lattice in two-dimensions is unstable to
the proliferation of dislocations in the presence of an
arbitrarily weak field of  random 
pinning centers\cite{N-S}$^,$\cite{zeng}.
Let us assume this to be  the case.
Let us also  assume that the interaction between dislocations can 
be neglected due to the screening action by the random pins\cite{fgl90}.
This requires a dilute concentration of dislocations in comparison
to the concentration of pinned vortices.
Consider then a {\it single} dislocation in the 2D vortex lattice
at zero temperature in the extreme type-II limit.
The latter implies  that the vortex
lattice is incompressible.  
The dislocation can therefore slide along
its glide plane\cite{book}, but  it cannot climb across it.
This would require the creation or the destruction of
vortices, which is prohibited by the extreme type-II limit.
Below, we shall demonstrate how randomly pinned vortices
in the 2D vortex lattice act to pin the dislocation itself along its
glide plane.

Consider a single dislocation  that can move along its glide plane
in the 2D vortex lattice with
randomly located material defects present.
Assume that a small fraction of the vortices are localized
at some subset of the pinning centers.  
The former is guaranteed at zero-temperature for a sparse array of
random pinning centers compared   to the density of vortices.
A vortex that lies at a point  $\vec R$ in the case of the
perfect triangular vortex lattice 
will in general be displaced to a position 
$\vec R + \vec u (\vec R)$  by the action of thermal
fluctuations and of the random pinning centers.
We shall now  make the approximation that the pinned vortices are
{\it fixed}:
\begin{equation}
\vec u (\vec R_i) = \vec v_i 
\quad {\rm for} \quad
i = 1,2, ..., N_{\rm pin},
\label{fix1}
\end{equation}
where $\vec R_i$ is the home site
of the vortex 
pinned down  at
$\vec R_i + \vec v_i$, and where $N_{\rm pin}$ denotes the total
number of pinned vortices. 
This approximation is valid for physics at large 
length scales compared to the
effective radius of a pinning center.
It then requires low magnetic fields
compared to the upper critical one
if the radius   of a pinning center is of order the coherence length.
The energy of the pinned vortex lattice 
is then given  by
%
\begin{equation}
E = {1\over 2} \mu_0\int d^2 R\, (\vec\nabla\times\vec u)^2
     + \int d^2 R\,  \vec\lambda\cdot  [\vec u - \vec v]
\label{elas1}
\end{equation}
in the continuum limit,
where $\mu_0$ denotes the shear modulus 
of the unpinned vortex lattice\cite{blatter}$^,$\cite{brandt},
and where
$\vec\lambda (\vec R) =
\sum_{i=1}^{N_{\rm pin}} \vec\lambda_i \, \delta^{(2)}(\vec R - \vec R_i)$
is the field of Lagrange multipliers
that is  introduced in order to
enforce each of the $N_{\rm pin}$ constraints (\ref{fix1}).
Also, the vortex lattice is  incompressible
in  the extreme type-II limit, 
and this requires that the displacement field satisfy the constraint
\begin{equation}
\vec\nabla\cdot\vec u = 0
\label{divzero}
\end{equation}
everywhere. 
By Eq. (\ref{elas1}),
the equilibrium configuration
$\vec u_0$  of the dislocation confined to its glide plane
then satisfies the field equation
\begin{equation}
\mu_0\vec\nabla\times\vec\nabla\times\vec u_0
     +  \vec\lambda_0 = 0
\label{fieldeq}
\end{equation}
everywhere. 
It can be used to show that
the elastic energy (\ref{elas1}) for a fluctuation about equilibrium,
$\vec u = \vec u_0 + \delta\vec u$,
takes the form
\begin{equation}
E = {1\over 2}\mu_0\int d^2 R\, (\vec\nabla\times\vec u_0)^2
          +{1\over 2}\mu_0\int d^2 R\, (\vec\nabla\times\delta\vec u)^2
          + \int d^2 R\, \delta\vec\lambda\cdot \delta\vec u,
\label{elas2}
\end{equation}
where 
$\delta\vec\lambda (\vec R) = 
\sum_{i=1}^{N_{\rm pin}}\, \delta\vec\lambda_i
\, \delta^{(2)} \, (\vec R - \vec R_i)$ 
is the field of the fluctuation
in the Lagrange multipliers,
$\delta\vec\lambda_i = \vec\lambda_i - \vec\lambda_i^{(0)}$.

To proceed further, it is convenient to decompose the displacement
of vortices into    {\it pure} wave and {\it pure} defect
components: $\vec u = \vec u_{\rm wv} + \vec u_{\rm df}$.
Suppose now that the dislocation is displaced by $\delta\vec R_{\rm df}$
along its glide plane with respect to its equilibrium position.
Notice then that the fluctuation in the defect component
corresponds to a pair of dislocations with equal and opposite
Burgers vectors oriented along the glide plane (see fig. \ref{diagram}):
\begin{equation}
\delta\vec u_{\rm df} (\vec R) =
 \vec u_{\rm df}^{(0)} (\vec R - \delta\vec R_{\rm df}) 
- \vec u_{\rm df}^{(0)} (\vec R),
\label{dudf}
\end{equation}
where $\vec u_{\rm df}^{(0)}(\vec R)$ denotes the displacement field of
the pure dislocation at its home site.
At  this stage it becomes important to observe that the pure wave and the
pure defect components do not interact elastically\cite{N-H}:
$\int d^2 R\, (\vec\nabla\times\vec u)^2 =
\int d^2R (\vec\nabla\times\vec u_{\rm wv})^2
+ \int d^2R (\vec\nabla\times\vec u_{\rm df})^2$.  
Application
of this fact to Eq. (\ref{elas2}) then ultimately yields the form
\begin{equation}
E =  E_{\rm df} + E_{\rm wv}^{(0)} 
          +{1\over 2}\mu_0\int d^2 R\, (\vec\nabla\times\delta\vec u_{\rm wv})^2
          + \int d^2 R\,  \delta\vec\lambda\cdot 
(\delta\vec u_{\rm wv} + \delta\vec u_{\rm df}),
\label{elas3}
\end{equation}
for the elastic energy, where
$E_{\rm wv}^{(0)} = 
(\mu_0 / 2)\int d^2 R\, (\vec\nabla\times\vec u_{\rm wv}^{(0)})^2$
is the wave contribution to the elastic energy at equilibrium.
Also,
$E_{\rm df} = (\mu_0 / 2) \int d^2 R\, (\vec\nabla\times\vec u_{\rm df})^2$
is the elastic energy of the displaced pure dislocation, which is constant.

The energy (\ref{elas3}) of the displaced dislocation
is therefore optimized by minimization 
with respect to the pure wave component $\delta\vec u_{\rm wv}$
along with the constraints 
\begin{equation}
\delta\vec u_{\rm wv} (\vec R_i) = -\delta\vec u_{\rm df} (\vec R_i) 
\quad {\rm for} \quad
i = 1, 2, ..., N_{\rm pin}.
\label{fix2}
\end{equation}
Its solution can be obtained by a straight forward generalization
of the solution for a pinned elastic string (see Appendix).
This yields 
\begin{equation}
\delta\vec u_{\rm wv} (\vec R) =
-\sum_{i = 1}^{N_{\rm pin}}\sum_{j = 1}^{N_{\rm pin}}
 \tensor G_{\perp} (\vec R - \vec R_i)\cdot
\tensor G_{i, j}^{-1}\cdot \delta\vec u_{\rm df} (\vec R_j),
\label{solution1}
\end{equation}
where
\begin{equation}
\tensor G_{\perp}(\vec R) = \sum_{\vec q} e^{i\vec q\cdot\vec R}
(\hat z\times\hat q) (\vec q L)^{-2} (\hat z\times\hat q)
\label{gperp}
\end{equation}
is the transverse Greens function 
over an $L\times L$ square region with periodic boundary conditions,
and where $\tensor G_{i, j}^{-1}$ is the inverse of the
$N_{\rm pin} \times N_{\rm pin}$ matrix
$\tensor G_{\perp}(\vec R_i - \vec R_j)$.  
Notice that (\ref{solution1}) manifestly satisfies 
the constraints (\ref{fix2}) and
the incompressibility requirement
$\vec\nabla\cdot\delta\vec u_{\rm wv} = 0$.
Also, direct substitution of the solution (\ref{solution1})
into Eq. (\ref{elas3}) yields a change  in  the elastic
energy due to the displacement of the dislocation equal to
\begin{equation}
\delta E_{\rm pin} = {1\over 2} \mu_0 \sum_{i=1}^{N_{\rm pin}} 
\sum_{j=1}^{N_{\rm pin}}
\delta\vec u_{\rm df}(\vec R_i)\cdot
 \tensor G_{i, j}^{-1}\cdot
\delta\vec u_{\rm df}(\vec R_j).
\label{elas4}
\end{equation}
Yet $\tensor G_{i, j}^{-1}$ is the inverse of the  2D Greens function,
$G = - \nabla^{-2}$,
projected onto transverse displacements (\ref{divzero}) and
onto the sites of the pinned vortices, $\{\vec R_i\}$.  
If 
these sites are {\it extensive and homogeneous}, 
then  they resolve unity at long wavelength:
$\sum_{i=1}^{N_{\rm pin}} |i\rangle\langle i|\cong 1$.
We therefore have that
$\tensor G_{i,j}^{-1} = 
\langle i|P_{\perp}^{-1} (-\nabla^2) P_{\perp}| j\rangle$
at longwavelength, where $P_{\perp}$ denotes the projection operator
for transverse displacements (\ref{divzero}).  
Substitution into Eq. (\ref{elas4}) then yields the expression
\begin{equation}
\delta E_{\rm pin} = {1\over 2} \mu_0 \int^{\,\prime} d^2 R\, 
(\vec\nabla\times\delta\vec u_{\rm df})^2
\label{elas5}
\end{equation}
for the change in the elastic energy due to the displacement of the dislocation,
where the prime notation signals that the integral 
has an ultraviolet cut-off ($R_{\rm pin}$)
 of order the average   spacing between
pinned vortices 
[see Eq. (\ref{special}) in the Appendix and ref. \cite{foot1}].
We conclude that the displacement of the dislocation along its glide
plane generates shear stress on the vortex lattice
via the array of pinned vortices (see fig. \ref{diagram}).
This then results in  a
restoring  Peach-Kohler force
on the displaced dislocation\cite{book}. 

Let us now compute the effective spring constant of the 
Peach-Kohler force experienced by the dislocation at small displacements
from equilibrium due to the array of pinned vortices:
\begin{equation}
\delta E_{\rm pin} = {1\over 2} k_{\rm pin} (\delta R_{\rm df})^2
\quad {\rm for}\quad
n_{\rm pin}    (\delta R_{\rm df})^2 \ll 1,
\label{hookes-pin}
\end{equation}
where $n_{\rm pin}$ denotes the density of pinned
vortices per layer.
The relative displacement field (\ref{dudf}) 
then corresponds to that of a {\it pure} dislocation pair
of extent $\delta R_{\rm df}$ that is  oriented along its glide plane.
Without loss of generality, it is sufficient to consider a pair of dislocations
centered at the origin, with the  glide plane
located  along the $x$-axis.
The displacement field is then  
given asymptotically by\cite{book}$^,$\cite{jpr01}
\begin{equation}
\delta\vec u_{\rm df} (\vec R)
 \cong  (b/\pi) (\delta R_{\rm df}) (X Y / R^4)\vec R,
\label{asymp}
\end{equation}
%
where  $b$ is one of the equal and opposite Burgers vectors
oriented  parallel to the glide vector $(\delta R_{\rm df}) \hat x$.
This expression is valid in the limit of small displacements
relative to the spacing between pinned vortices: 
$n_{\rm pin} (\delta R_{\rm df})^2 \ll 1$.
Substitution into expression (\ref{elas5}) for the  change in the elastic
energy yields
the result 
\begin{equation}
k_{\rm pin} = (2 n_{\rm pin} \pi R_{\rm pin}^2)^{-1}
(\mu_0 b^2) n_{\rm pin}
\label{kpin}
\end{equation}
for the effective spring constant of the Peach-Kohler force (\ref{hookes-pin}),
where $R_{\rm pin}$ denotes the natural ultraviolet cutoff of
the array of pinned vortices\cite{foot1}:
$n_{\rm pin}\cdot \pi R_{\rm pin}^2\sim 1$.
Equations (\ref{hookes-pin}) and (\ref{kpin}) represent the final
result of this section.  It indicates that the incompressible vortex
lattice confined to two dimensions does {\it not} respond plastically to 
small shear stress\cite{book}
when a  dilute enough  concentration of unbound dislocations
are quenched in.
Instead,  the response to small
shear stress should remain elastic, 
like in the pristine case\cite{blatter}$^,$\cite{brandt},
due to the pinning of the quenched-in dislocations.

\section{Uniformly Frustrated $XY$ Model}
\label{ufxy}

The minimal description of the mixed phase in a layered superconductor 
is given by a stack of isolated
$XY$ models with uniform frustration
over the square lattice\cite{hattel_xy}.
Both the effects of magnetic screening and of
Josephson coupling between layers are 
neglected in this approximation.
The thermodynamics of each layer is then 
determined by the superfluid kinetic energy
\begin{equation}
 E_{XY}^{(2)} = -\sum_{\mu = x, y} \sum_r  J_{\mu} {\rm cos}
[\Delta_{\mu}\phi  - A_{\mu}]|_r ,
\label{2dxy}
\end{equation}
which is a  functional of the superconducting
phase $\phi(r)$ over the square lattice.
The local phase rigidities within layers, $J_{x}$ and $J_{y}$,
are assumed to be constant over most of the
nearest-neighbor links,
with the exception of those links in the vicinity of a pinning site.
The   vector potential
$A_{\mu} = (0, 2\pi f x/a)$
represents the magnetic induction,
$B_{\perp} = \Phi_0 f / a^2$,
oriented perpendicular to each layer.
Here $a$ denotes the square lattice constant for each layer,
which is of order
the zero-temperature coherence length.
Also,  $\Phi_0$ denotes the flux quantum, 
and $f$ denotes the concentration of planar vortices per site.
After taking  the Villain approximation,
which is generally valid at low temperature\cite{villain},
a series of standard manipulations then lead to
a Coulomb gas ensemble with pinning centers    that
describes   the vortex degrees of
freedom on the dual square lattice\cite{cec03}.
The ensemble for each layer is weighted by the Boltzmann
distribution set by the energy functional
\begin{equation}
  E_{\rm vx}  =  (2\pi)^2   \sum_{(\vec R, \vec R^{\prime})}
\delta Q\ J_0 G^{(2)}
\ \delta Q^{\prime}
+  \sum_{\vec R} V_{\rm pin}  \, | Q  |^2
 \ ,
\label{evx}
\end{equation}
written in terms of the integer vorticity field $Q (\vec R)$ over
the sites $\vec R$ of the dual lattice in that  layer,
and of the fluctuation $\delta Q = Q - f$.
A logarithmic interaction, $G^{(2)} = - \nabla^{-2}$,
exists between the vortices, with a strength $J_0$ equal
to the gaussian  phase rigidity.
Last, $V_{\rm pin} (\vec R)$
is the resulting pinning potential\cite{cec03}.

The 2D Coulomb gas ensemble (\ref{evx}) can be used to
test for the presence or the absence of  superconductivity.
In particular,
the macroscopic phase rigidity parallel to the layers
is  given by
one over its dielectric constant\cite{eps}:
%
\begin{equation}
{\rho_s^{(2D)}  /    {J_0}}    =
 1 - \lim_{k \rightarrow 0} (2\pi / \eta_{\rm sw})
\langle \delta Q_{\vec k} \delta Q_{-\vec k} \rangle
 / k^2 a^2 {\cal N_{\parallel}}  \ .
\label{epsinv}
\end{equation}
Here  $\delta Q_{\vec k} =  Q_{\vec k} - \langle Q_{\vec k}\rangle$
is the fluctuation in the Fourier transform of the vorticity:
$Q_{\vec k} = \sum_{\vec R} Q (\vec R) e^{i \vec k\cdot \vec R}$.
 Also,
$\eta_{\rm sw} = k_B T / 2\pi J_0$
is the  spin-wave component of the
phase-correlation exponent,
and
$\cal N_{\parallel}$ denotes the number of points in the square-lattice grid.
Now suppose that a given vortex is displaced
by $\delta\vec u$ with respect to its equilibrium
location at zero temperature, $\vec u_0$.
Conservation of vorticity dictates that
the fluctuation in the vortex number is given by
$\delta Q = - \vec\nabla\cdot\delta\vec u$.
Substitution
into Eq. (\ref{epsinv}) then yields the result\cite{jpr04a}
%
\begin{equation}
\rho_s^{(2D)} / J_0 = 1 - ( \eta_{\rm vx}^{\prime} / \eta_{\rm sw} )
\label{rho2d1}
\end{equation}
for the phase rigidity in terms of the
vortex component of the phase-correlation exponent,
\begin{equation}
\eta_{\rm vx}^{\prime} = \pi \Bigl\langle
\Bigl[\sum_{\vec R}^{\qquad\prime} \delta\vec u \Bigr]^2\Bigr\rangle /
N_{\rm vx} a_{\rm vx}^2.
\label{etavx1}
\end{equation}
The latter monitors fluctuations of the
center of mass of the vortex lattice\cite{jpr01}.
Above, $N_{\rm vx}$ denotes the number of vortices,
while $a_{\rm vx} = a / f^{1/2}$
is equal to the square root of the area per vortex.
Also, the prime notation above signals that the summation
is restricted to the vortex lattice.
To proceed further,
we again express
the displacement field  as a superposition of {\it pure}
wave and of  {\it pure} defect components of the triangular
vortex lattice\cite{jpr01}:
$\delta\vec u = \delta\vec u_{\rm wv} + \delta\vec u_{\rm df}$.
Observe now that $\sum_{\vec R}^{\prime} \delta\vec u_{\rm wv} = 0$ 
under periodic boundary conditions
if rigid translations of the 2D vortex lattice are not possible.
The latter
is  achieved by the array of pinned vortices (\ref{fix1})
through the elastic forces (\ref{elas1}).
By Eq. (\ref{etavx1}), 
we therefore have the result
\begin{equation}
\eta_{\rm vx}^{\prime} = \pi \Bigl\langle
\Bigl[\sum_{\vec R}^{\qquad\prime}
 \delta\vec u_{\rm df} \Bigr]^2\Bigr\rangle / N_{\rm vx} a_{\rm vx}^2
\label{etavx2}
\end{equation}
for the fluctuation in the center of mass of the 2D vortex lattice.
The  degree of phase coherence in the pinned vortex lattice 
is therefore insensitive to its {\it pure} wave contribution.

Consider now the hexatic vortex glass\cite{hexglass}$^,$\cite{chudnovsky},
with a collection of $N_{\rm df}$ randomly located unbound dislocations that
are quenched in by the random array of pinned vortices\cite{N-S}$^,$\cite{zeng}.
Suppose also that the temperature is low enough so that
the thermal excitation of  pairs of dislocations in the 
vortex lattice can be neglected.
Within the elastic medium  description ({\ref{elas1}),
the  pure defect component of the net displacement field 
is just  a simple sum of  the displacements due to each individual dislocation.
And by analogy with the hexatic liquid
phase of the pure 2D vortex lattice\cite{N-H},
we shall assume that interactions in between the unbound dislocations
can be neglected,
be they direct or be they transmitted through the field of pinned vortices.
Expression (\ref{etavx2}) for
the fluctuation of the center of mass of the 2D vortex lattice
then reduces to
\begin{equation}
\eta_{\rm vx}^{\prime} \cong \pi 
\overline{\Bigl\langle
\Bigl[\sum_{\vec R}^{\qquad\prime} 
\delta\vec u_{\rm df}^{(1)} \Bigr]^2
\Bigr\rangle} n_{\rm df},
\label{etavx3}
\end{equation}
where $\delta\vec u_{\rm df}^{(1)} (\vec R)$ denotes the fluctuation field of 
a given dislocation displaced along its glide plane (\ref{dudf}),
where $n_{\rm df} = N_{\rm df}/N_{\rm vx} a_{\rm vx}^2$
is the density of unbound dislocations per layer, 
and where the overbar notation denotes a bulk average.
A lone dislocation
roams along its glide plane
to an extent that
is  vanishingly small, however, in the zero-temperature limit:
$\delta\vec R_{\rm df}\rightarrow 0$ as $T\rightarrow 0$.
Without loss of generality, we can  then use  the asymptotic expression
(\ref{asymp}) for the corresponding fluctuation in the displacement field,
$\delta\vec u_{\rm df}^{(1)}$.
The fluctuation in the center of mass of the
2D vortex lattice (\ref{etavx3}) is 
dominated by the ``diagonal'' on-site contribution,
$\overline{\langle
[\sum_{\vec R}^{\quad\prime}
|\delta\vec u_{\rm df}^{(1)}|^2\rangle}$,
which yields the estimate\cite{jpr04a}
\begin{equation}
\eta_{\rm vx}^{\prime}  \cong 
   n_{\rm df}
\overline {\langle |\delta  R_{\rm df} |^2 \rangle}
(b / 2 a_{\rm vx})^2 {\rm ln}\, R_0 / a_{\rm df}.
\label{etavx4}
\end{equation}
%
Here $a_{\rm df}$ is the core diameter of a   dislocation,  while
$R_0$ is an infrared cut-off.  
The above logarithmic
divergence associated with the latter scale justifies the neglect 
of the contribution
to the fluctuation in the center of mass (\ref{etavx3})
by   the autocorrelator
$\overline{\langle \delta \vec u_{\rm df}^{(1)}
 \cdot \delta\vec u_{\rm df}^{(1)\, \prime} \rangle}$
at different points.
This is due to the fact ({\it i}) that
$-\overline{\langle \delta\vec u_{\rm df}^{(1)} (a) 
\cdot \delta\vec u_{\rm df}^{(1)} (b)\rangle}$
must decay faster than
$[\overline{\langle |\delta\vec u_{\rm df}^{(1)}|^2 \rangle}
/2\pi \, {\rm ln} (  R_0 / a_{\rm df})]
 (a_{\rm vx}/R_{ab})^2$
by Eq. (\ref{etavx3})
because $\eta_{\rm vx}^{\prime} > 0$, and
to the fact ({\it ii}) that the former autocorrelator is short range
as a result of disordering by the quenched-in dislocations.
Last, given that expression (\ref{etavx4}) 
was obtained by  neglecting interactions in  between isolated dislocations,
it is natural to assume that the infrared scale $R_0$ that appears 
there is set by their density, $n_{\rm df}$.

\section{Low-temperature Phase Coherence}
\label{pc}
We shall now assemble the results of the previous two sections and
compute the macroscopic phase rigidity of a defective vortex lattice
in two dimensions.
Recall  that
the energy cost for a small displacement of the dislocation
along its glide plane takes the form
$\delta E_{\rm df} = {1\over 2} k_{\rm pin} |\delta R_{\rm df}|^2$,
where $k_{\rm pin}$ is the effective spring constant
(\ref{kpin}) due to the randomly pinned vortices.
This approximation for the elastic energy of the dislocation along
its glide plane is valid in the zero-temperature limit, $T\rightarrow 0$,
where it amounts to a saddle-point approximation for the Boltzmann
weight in the thermal average $\langle |\delta R_{\rm df}|^2\rangle$.
The periodic Peierls-Nabarro potential energy along the
glide plane of the dislocation shall be neglected for the moment\cite{book}.
Application of the equipartition theorem to expression (\ref{etavx4}) 
for the fluctuation of the center of mass of the 2D vortex lattice
then yields the result
\begin{equation}
\eta_{\rm vx}^{\prime}  \cong
   (k_B T) (n_{\rm df}/ k_{\rm pin})
(b / 2 a_{\rm vx})^2 {\rm ln}\, R_0 / a_{\rm df},
\label{etavx5}
\end{equation}
which notably vanishes linearly with temperature.
Substitution into expression (\ref{rho2d1}) in turn yields the  result
\begin{equation}
\rho_s^{(2D)}  / J_0 \cong  1 -
   (2\pi J_0) (n_{\rm df}/ k_{\rm pin})
(b / 2 a_{\rm vx})^2 {\rm ln}\, R_0 / a_{\rm df}
\label{rho2d2}
\end{equation}
for the 2D phase rigidity in the zero-temperature limit.
Substitution of the result
(\ref{kpin}) for the   effective spring constant
above
yields the final formula for the macroscopic 2D phase rigidity 
near zero temperature,
\begin{equation}
{\rho_s^{(2D)}  \over {J_0}}
 \cong  1 -
(n_{\rm pin}\cdot \pi R_{\rm pin}^2)
{\pi J_0\over{\mu_0 a_{\rm vx}^2}}
{N_{\rm df}\over{N_{\rm pin}}}
 {\rm ln}\,\Biggl({R_0 \over{ a_{\rm df}}}\Biggr),
\label{rho2d3}
\end{equation}
in terms of the number of  unbound dislocations, $N_{\rm df}$,
the number of vortices, $N_{\rm vx}$, and of the number
of pinned vortices, $N_{\rm pin}$, in an isolated 2D vortex lattice.
Recall now the estimate
for the shear modulus of the unpinned vortex lattice 
in the extreme type-II limit\cite{blatter}$^,$\cite{brandt},
$\mu_0 a_{\rm vx}^2 = (\pi/4) J_0$.
Substitution into expression (\ref{rho2d3})
for the superfluid density then yields the
yet simpler result
\begin{equation}
{\rho_s^{(2D)}  \over {J_0}}
 \cong  1 -
(2n_{\rm pin}\cdot \pi R_{\rm pin}^2)
{N_{\rm df}\over{N_{\rm pin}}}
 {\rm ln}\,\Biggl({R_0 \over{ a_{\rm df}}}\Biggr)^2
\quad {\rm near} \quad T=0.
\label{rho2d4}
\end{equation}
The effect of the Peierls-Nabarro potential energy
with period $\vec b$
that any dislocation experiences along its glide plane 
has been neglected above\cite{book}.
It becomes useful in this instance to define the temperature scale
\begin{equation}
k_B T_0 = k_{\rm pin} b^2 \sim (\mu_0 b^2) (n_{\rm pin} b^2),
\label{T_0}
\end{equation}
at which point thermally induced excursions of the dislocation
about its home site are typically of the size of  a Burger's vector, $b$.
Notice first that $k_B T_0$ is an  extremely small fraction
of  the
elastic energy scale $\mu_0 b^2$ if the  concentration of pinned
vortices is dilute.
Typical excursions of the dislocation will be large compared to
the Burgers vector at temperatures above this extremely low
scale:  on average, $|\delta R_{\rm df}| > b$ at  $T > T_0$. 
In such case,
the periodic  Peierls-Nabarro potential can be neglected 
because the thermal motion of
the dislocation becomes insensitive to its
relatively short period $b$.
The periodic potential will take effect at  extremely low
temperature $T\ll T_0$, on the other hand, in which case it
will tend to localize the dislocation even further about its home site.

Let us finally close the chain of calculations
by estimating the ratio of  the number
of unbound dislocations that are quenched into
the vortex lattice to   the  number of
pinned vortices.
Notice that it determines the phase rigidity (\ref{rho2d4})
of the defective vortex lattice
at $T > T_0$.  A variational calculation by Mullock and Evetts
finds that this ratio is given by
\begin{equation}
{N_{\rm df}\over{N_{\rm pin}}} = 
\Biggl[{\pi\over{
{\rm ln} (n_{\rm df} a_{\rm df}^{\prime 2})^{-1}}}\Biggr]^2
\Biggl({f_{\rm pin}\over{\mu_0 b}}\Biggr)^2,
\label{ratio}
\end{equation}
where $f_{\rm pin}$ denotes the maximum pinning force,
and where $ a_{\rm df}^{\prime}$ is of order the core diameter
of a dislocation in the vortex lattice\cite{M-E}.
This result is valid only 
in the collective-pinning regime
at $N_{\rm df}\leq N_{\rm pin}$.
Equation (\ref{ratio})
therefore implies a small ratio of topological
defects to pinned vortices, $N_{\rm df}\ll N_{\rm pin}$,
for weak pinning forces compared to the elastic forces, 
$f_{\rm pin}\ll \mu_0 b$.  
Substitution of the estimate for the shear modulus 
quoted earlier\cite{blatter}$^,$\cite{brandt}
yields the field dependence 
$N_{\rm df} / N_{\rm pin} = B_{\rm cp} / B$
for this ratio (\ref{ratio}),
where
$B_{\rm cp} = 
(\sqrt{3} / 2)
[\pi/{\rm ln} (n_{\rm df} a_{\rm df}^{\prime 2})^{-1}]^2
(4 f _{\rm p} / \pi J_{0})^2 \Phi_0$
is the threshold  magnetic field above which collective pinning holds.
Last,
the effect of substrate pinning by the $XY$ model grid (\ref{2dxy}),
which tends to suppress the number of 
unbound dislocations even further,
is neglected here.
This is valid in the regime of dilute vortex lattices
compared to the model grid\cite{hattel_xy}, $f <  1/36$.

Three important conclusions can be reached from the  
estimate for the 
degree of macroscopic phase coherence encoded by
Eqs. (\ref{rho2d4}) and (\ref{ratio}) above. 
First, observe that $N_{\rm vx}$, $N_{\rm pin}$ and $N_{\rm df}$
are all extensive thermodynamic variables that scale
with the area of each layer, $L^2$.
The macroscopic
phase rigidity (\ref{rho2d4}) hence attains its maximum value $J_0$
near   zero temperature 
when the total number of unbound dislocations is subthermodynamic:
e.g. if  $N_{\rm df}\propto L$, or if  $N_{\rm df}$ remains
finite as $L\rightarrow\infty$.
This  state   is then
a Bragg glass\cite{N-S}$^,$\cite{G-LeD}.
The variational result (\ref{ratio}) for
the number of unbound dislocations
obtained by Mullock and Evetts\cite{M-E}
indicates that it
exists only in the absence of bulk
point pins.
Other types of pinning, such as surface barriers or planar defects,
must therefore be present in order to impede flux flow
by the 2D Bragg glass\cite{jpr01}.
Second,  recall that 
$n_{\rm pin}\cdot \pi R_{\rm pin}^2\sim 1$.
Expression (\ref{rho2d4}) therefore implies that 
weaker macroscopic phase coherence
exists near    zero temperature
at   dilute concentrations of unbound dislocations:
$\rho_s^{(2D)} (0+) > 0$ for $N_{\rm df} \ll N_{\rm pin}$.
Such a state is then a hexatic vortex glass\cite{hexglass}$^,$\cite{chudnovsky}.
The variational result (\ref{ratio}) for
the number of unbound dislocations
indicates that this state exists at weak pinning,
$f_{\rm pin}\ll\mu_0 b$, 
which occurs at large magnetic fields
$B\gg B_{\rm cp}$.
Third, expression (\ref{rho2d4}) also implies   that a pinned vortex liquid 
that  shows {\it no} macroscopic phase coherence is possible
in the zero-temperature limit
at sufficiently high concentrations of unbound dislocations:
$\rho_s^{(2D)} (0+) = 0$ if $N_{\rm df}\sim N_{\rm pin}$.  
This phase is then 
a (pinned) hexatic vortex liquid\cite{N-H}$^,$\cite{jpr01}.
The variational result ({\ref{ratio}) for the number of dislocations
indicates that such a phase-incoherent state 
can occur in the regime of strong pinning forces, at low fields
compared to the collective pinning threshold $B_{\rm cp}$.
We remind the reader that expression (\ref{rho2d4}) was derived by
neglecting the interactions in between dislocations.
This approximation, and Eq. ({\ref{rho2d4}) as a result,
may not necessarily be valid in the regime of 
relatively dense dislocations
last discussed.


\section{Discussion}

The results of the previous sections are summarized by Table \ref{proper}.
Here  the low-temperature phases are listed by 
increasing order of the random
pinning force, $f_{\rm pin}$.
Below, we confront Table \ref{proper} 
with previous theoretical  work on 2D vortex matter.

{\it Order Parameters.}
It is natural to ask what order parameters characterize the phases
listed in Table \ref{proper}.
Let us begin by defining the auto-correlation function
\begin{equation}
G_{BrG} (\vec r) = \overline {\langle{\rm exp}\, i [\phi({\vec r}^{\,\prime})
-\phi(\vec r + {\vec r}^{\,\prime}) 
+\int_{\vec r^{\,\prime}}^{\vec r + {\vec r}^{\,\prime}} 
d\vec l\cdot \vec A / a]\rangle}
\label{g_bg}
\end{equation}
for the  Bragg-glass order parameter,
where the overbar denotes a bulk average over $\vec r^{\,\prime}$.
Drawing the analogy with the thermal degradation
of  phase coherence 
in the pristine 2D vortex lattice\cite{jpr01} implies
that  $G_{BrG}(\vec r)$ is {\it not} short range in the zero-temperature
limit if no  unbound dislocations are quenched in.  
The  latter is consistent with with the properties of the Bragg glass
listed in Table \ref{proper} (see ref. \cite{G-LeD}).  
We conclude that the Bragg-glass  phase
studied here displays conventional phase coherence at long range.
Again 
in analogy with the case of thermal disordering
of the pristine vortex lattice\cite{jpr01}, 
the unbound dislocations that are quenched inside of the hexatic vortex glass,
on the other hand,
will result in short-range order in $G_{BrG}(\vec r)$ over a scale
set by the density of such defects, $n_{\rm df}$.
The absence of
Bragg-glass order in the   hexatic vortex glass,
as defined above by Eq. (\ref{g_bg}), is then consistent with the presence of
unbound dislocations in the vortex lattice\cite{G-LeD}.

Following Fisher, Fisher and Huse\cite{ffh},
we can next define the vortex glass auto-correlation function
\begin{equation}
G_{VG} (\vec r) = \overline {|\langle{\rm exp}\, i [\phi({\vec r}^{\,\prime})
-\phi(\vec r + {\vec r}^{\,\prime})]\rangle|^2}.
\label{g_vg}
\end{equation}
In the zero-temperature limit,
this function is notably identical to unity if the groundstate configuration,
$e^{i\phi_0}$,
is unique.
Recall now that the present treatment
of the  hexatic vortex glass has been restricted to 
the limit of vanishing dislocation density,
in which case a unique groundstate is to be expected.
The
auto-correlation
function $G_{VG}(\vec r)$ then is {\it not} short range
for the case of the hexatic vortex glass in the
limit of weak disorder pinning.
The zero-temperature configuration may not be unique,
on the other hand, 
for the case of the (pinned) hexatic vortex liquid.  
The number of unbound dislocations is comparable to the number of pinned
vortices in such case.   Interactions in between dislocations
may become important,
and these  may frustrate the confining action
on the dislocations
by the pinned vortices. 
The end result could be multiple groundstates  that lead to a
vortex glass auto-correlation function $G_{VG}(\vec r)$
that shows only  short-range order.

Consider again the (pinned) hexatic vortex liquid state that is
possible  
at relatively dense concentrations of unbound
dislocations, $N_{\rm df}\sim N_{\rm pin}$,
by expression (\ref{rho2d4}) 
for the superfluid density.
Some fraction of the
pairs of five-fold and seven-fold coordinated disclinations that
make up the unbound dislocations present in this state
will unbind in the strong pinning limit, $f_{\rm pin} \gg \mu_0 b$.
This is confirmed by direct Monte Carlo simulations\cite{jpr-cec05}
of the Coulomb gas ensemble (\ref{evx}).  
Such a state then shows only short-range translational and orientational order.
It also cannot have a net superfluid density,
 $\rho_s^{(2D)} > 0$,
since it is yet more disordered than the hexatic vortex liquid from
whence it originates. 
For the same reason,
it can neither show long-range Bragg glass nor vortex glass order
in the respective  auto-correlation functions (\ref{g_bg}) and (\ref{g_vg}).
This strongly pinned state is then a
``conventional''
 vortex liquid (see Table \ref{proper}). 

{\it Vortex Glass in 2D?}
Fisher, Fisher and Huse argue
in ref. \cite{ffh} that the 
vortex-glass state is not possible in two dimensions above
zero temperature.  This statement conflicts at first sight
with the phase-coherent  vortex lattice state with
a dilute concentration of quenched-in  dislocations that we have
discovered in the uniformly frustrated 
$XY$ model (\ref{2dxy}) in the limit of
weak random  pinning. 
This state also shows vortex-glass order (\ref{g_vg})!
Study of ref. \cite{ffh} reveals that the authors
presume the strong-pinning limit, however.  The vortex glass is necessarily
amorphous under such conditions,
where it possesses a net concentration of unbound disclinations.
It hence lies in the same  topological class as the 
``conventional''
vortex liquid state discussed above and listed in  Table \ref{proper}.  
No conflict  then truly exists between ref. \cite{ffh} and
the present results concerning the impossibility of observing an
amorphous vortex glass in two dimensions.
The latter is topologically distinct from the hexatic vortex glass
discovered here.

The hexatic vortex glass will also have some concentration of bound
pairs of dislocations quenched into the vortex lattice in the
zero-temperature limit.  Vinokur and co-workers have argued
that the absence of infinite potential barriers along the corresponding
glide planes will result in  thermally activated plastic creep
of magnetic flux
due to the diffusion of such pairs of dislocations,
and in  thermally activated electrical resistance 
as a result\cite{vino90}$^,$\cite{blatter}.
Hence, although a  stack of uncoupled sheets of hexatic vortex
glass shows magnetic screening in direct proportion to the superfluid
density, $\rho_s^{(2D)}$, this system may not in fact be a perfect conductor!
Flux creep immediately becomes neutralized,
however,
once macroscopically 
big layers
(compared to the Josephson penetration depth)
are coupled through the Josephson effect.
This is
due to the binding of pairs of dislocations
into ``quartets'' that carry no net magnetic
flux\cite{fgl90}$^,$\cite{blatter}.  
The possibility just raised 
of three-dimensional  vortex matter that is  ohmic, but that nevertheless
shows  macroscopic  phase coherence,  therefore remains unrealistic.

\section{Conclusions}
\label {conc}

We have demonstrated that incompressible vortex lattices
that are confined to two dimensions
and  that are void of unbound  disclinations
show macroscopic phase coherence near zero temperature in the limit
of weak random pinning.  The latter ensures that the total number of unbound
dislocations quenched into the vortex lattice by the randomly
pinned vortices is small in comparison to the
total number of  such pins.
This in turn  results in a  net superfluid density.
The hexatic vortex glass predicted here is consistent
with the observation of  isolated dislocations that are  quenched into the
vortex lattice of extremely layered high-temperature superconductors
at low temperature\cite{hexglass}, and with the observation
of superconductivity in 2D Josephson junction arrays in external magnetic
field\cite{arrays}.
How exactly  such a superconducting vortex lattice transits
into a vortex liquid with increasing temperature remains unclear.   
Equations (\ref{rho2d1}) and (\ref{etavx2}) indicate that
the superfluid density vanishes either
once  the quenched-in dislocations
delocalize and begin to cross the length of the vortex lattice,
or once thermally activated pairs of dislocations
unbind and begin to cross the length of the vortex lattice\cite{jpr01}.
Both mechanisms cause  plastic creep of the 2D vortex lattice\cite{book},
which destroys macroscopic phase coherence.
Continuity implies that the melting
temperature $T_g^{(2D)}$  of the defective  vortex lattice
lies near that of the 
pristine vortex lattice\cite{hattel_xy}, $k_B T_m^{(2D)}\cong J_0/20$,
in the limit of weak random pinning.


\acknowledgments
The author thanks M. Maley and L. Civale for useful discussions.
He is also grateful for the hospitality of the 
Superconductivity Technology Center at Los Alamos National Laboratory
during the last stages of this work.

\appendix*\section{Pinned Elastic String}
\label{apndx}
Consider a tense elastic string of length $L$ that lies along the
$x$ axis under periodic boundary conditions, and suppose that only
transverse displacements, $u(x)$,  along the $y$ axis are allowed. 
Suppose further that the string is pinned down at $N_{\rm pin}$ sites:
\begin{equation}
u(x_i) = v_i \quad {\rm for} \quad i = 1,2, ..., N_{\rm pin}.
\label{ueqv}
\end{equation}
The shape of the string with the lowest energy can then be
determined by minimizing the
elastic energy along with appropriate terms that enforce the
constraints:
\begin{equation}
E = {1\over 2} \mu_0\int_0^L dx \Biggl({d u\over d x}\Biggr)^2
     + \int_0^L dx\,  \lambda\cdot  [u - v].
\label{elas}
\end{equation}
Here  $\mu_0$ denotes the shear modulus, while
the field
$\lambda (x) = \sum_{i=1}^{N_{\rm pin}} \lambda_i\delta (x-x_i)$
is  weighted by the 
Lagrange  multipliers  $\lambda_i$ that correspond to 
each of the constraints (\ref{ueqv}).  
The configuration  that minimizes the elastic energy (\ref{elas})
satisfies the field equation
\begin{equation}
-\mu_0 {d^2 u\over{d x^2}} + \lambda = 0
\label{field}
\end{equation}
everywhere.
Equation (\ref{elas}) can be easily minimized in
the wave representation,
\begin{equation}
u(x) = \sum_q u_q e^{iqx}
\quad {\rm and} \quad
\lambda(x) = \sum_q \lambda_q e^{iqx},
\label{fourier}
\end{equation}
%
expressed as a sum over
allowed wavenumbers $q$ that are multiples of $\pm 2\pi/L$.
One then  obtains the solution
\begin{equation}
u (x) = \sum_{i=1}^{N_{\rm pin}}\sum_{j=1}^{N_{\rm pin}}
       G(x-x_i)\, G_{i,j}^{-1}\, v_j ,
\label{solution2}
\end{equation}
where $G(x) = \sum_q e^{iqx} / L q^2$ is the Greens functions in 
one dimension, and where $G_{i,j}^{-1}$ is the inverse of the
$N_{\rm pin} \times N_{\rm pin}$ matrix $G(x_i - x_j)$.
Notice that the above solution satisfies the constraints (\ref{ueqv}).
It also satisfies the field equation (\ref{field}),
with Lagrange multipliers 
$\lambda_i = - \mu_0 \sum_{j=1}^{N_{\rm pin}} G_{i,j}^{-1} v_j$
that weight the field $\lambda (x)$
at each of the pins located at $x_i$.
Direct substitution of the solution (\ref{solution2}) into the elastic energy
(\ref{elas}) then  yields the result
\begin{equation}
E = (\mu_0/ 2) \sum_{i=1}^{N_{\rm pin}}\sum_{j=1}^{N_{\rm pin}}
    v_i\, G_{i,j}^{-1}\, v_j.
\label{result}
\end{equation}
It reduces to  the expression
\begin{equation}
E_{1,2} = (\mu_0 / 2) (v_1 - v_2)^2 /|x_1 - x_2| 
\label{special}
\end{equation}
for the elastic energy
in the special case of two pins at the
thermodynamic limit, $L\rightarrow\infty$.  
Here we used the result 
$G(x) = G(0) - (|x|/2)$
for the  Greens function in one dimension, 
with a limiting value  $G(0)\rightarrow\infty$ for the constant.

\begin{figure}
\includegraphics[scale=0.36, angle=-90]{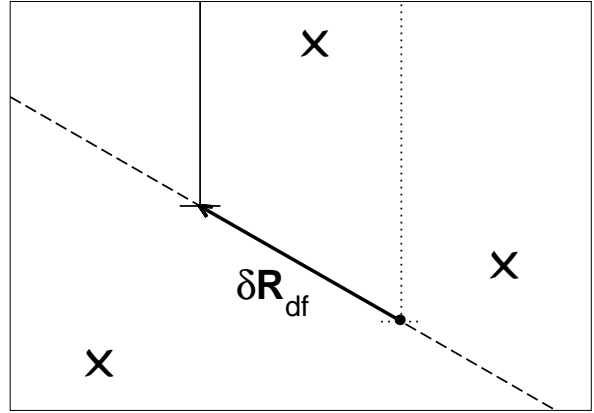}
\caption{Shown is a diagram for glide by a dislocation (upside-down ``T'')
in the presence of randomly pinned vortices (``X'').}
\label{diagram}
\end{figure}  


\bigskip\bigskip\bigskip

\begin{table}
\begin{center}
\begin{tabular}{|c|c|c|c|c|}
\hline
 disorder index & phase & $\rho_s^{(2D)} (0+) /J_0$ 
& unbound dislocations? & unbound disclinations? \\
\hline
1 & Bragg glass & unity & no & no \\
2 & hexatic vortex glass & fraction  & yes & no  \\
3 & hexatic vortex liquid & zero & yes  & no \\
4 & vortex liquid &  zero & yes & yes \\
\hline
\end{tabular}
\caption{Listed are physical properties that characterize
the low-temperature phases of vortex matter in two dimensions 
at the extreme type-II limit.
The list  is  indexed by increasing levels of disorder.}
\label{proper}
\end{center}
\end{table}


\begin{thebibliography}{}
\bibitem{foot1}
Arrangements of pinned vortices that show clusters and voids do not possess
a well defined ultraviolet scale, $R_{\rm pin}$.
They can occur near the lower-critical magnetic field,
in soft vortex lattices with a purely random arrangement
of columnar  pinning centers,
at relatively low  matching fields
(see  ref. \cite{menghini}).
\end{thebibliography}
\end{document}